\documentclass[11pt]{article} 
\usepackage[paperheight=9.44in,paperwidth=6.69in,left=1.85cm,right=1.8cm,bottom=2.5cm,top=3cm,twoside]{geometry}

\usepackage[linesnumbered,ruled,vlined,boxed]{algorithm2e}
\usepackage{authblk} 
\usepackage{fancyhdr} 
\usepackage{lastpage} 
\usepackage{textcomp} 
\usepackage{amsmath} 
\usepackage{amssymb} 
\usepackage{amsthm} 
\usepackage{graphicx} 
\usepackage{multirow} 
\usepackage{hyperref} 
\usepackage{longtable} 
\usepackage{float} 

\newcommand{\linia}{\noindent\rule{\linewidth}{0.5mm}\hrulefill} 

\SetAlFnt{\small}
\SetAlCapFnt{\small}
\SetAlCapNameFnt{\small}
\SetAlCapHSkip{0pt}
\IncMargin{-\parindent}

\usepackage{times}
\usepackage{titlesec}

\titleformat*{\section}{\large\bfseries}
\titleformat*{\subsection}{\normalsize\bfseries}

\fancypagestyle{firststyle}
{
	\fancyhf{}
	\fancyfoot{}
	\fancyhead{}
	\fancyhead[CE]{\upshape Zeszyty Naukowe WWSI, No 16, Vol. 11, 2017, pp. \thepage-\pageref{LastPage}}
	\fancyhead[CO]{\upshape Zeszyty Naukowe WWSI, No 16, Vol. 11, 2017, pp. \thepage-\pageref{LastPage}}
	\fancyfoot[L,R]{\footnotesize\bfseries Manuscript received June 3, 2017}
	\fancyfoot[LE,RO]{\thepage}
	}

\title{\large \bfseries New Interpretation of Principal Components Analysis} 
\author{\normalsize Zenon Gniazdowski\thanks{E-mail: zgniazdowski@wwsi.edu.pl}}
\affil{\normalsize Warsaw School of Computer Science}

\date{\vspace{-5ex}}
\providecommand{\keywords}[1]{\textbf{\textit{Keywords ---}} #1}

\begin{document}
	
	\maketitle 
	\thispagestyle{firststyle} 
	
	\linia
	\begin{abstract}
		\noindent A new look on the principal component analysis has been presented. Firstly, a geometric interpretation of determination coefficient was shown. 
		In turn, the ability to represent the analyzed data and their interdependencies in the form of easy-to-understand basic geometric structures was shown.
		As a result of the analysis of these structures it was proposed to enrich the classical PCA. In particular, it was proposed a new criterion for the selection of important principal components and a new algorithm for clustering primary variables by their level of similarity to the principal components. Virtual and real data spaces, as well as tensor operations on data, have also been identified.The anisotropy of the data was identified too.
	\end{abstract}
	
	\section{Introduction}
	In the method of principal component, a primary set of data consisting of $n$ mutually correlated random variables can be represented by a set of independent hypothetical variables called principal components. 
	A new dataset typically contains fewer variables than the original data. 
	The smaller number of principal components contains almost the same information as the full set of primary variables \cite{hand2001}\cite{Larose2006DMmethods}.
	
	This work is an attempt to make a new interpretation of results of the classical principal components analysis. Here it should be noted that it is not the purpose of this work full presentation of the principal components method. This can be found in the literature. So, tips on how to formulate a problem in the form of an optimization task can be found in \cite{hand2001}. Examples of the use of this method can be found in \cite{hand2001} and \cite{Larose2006DMmethods}. In this article, the principal component method will be presented in such a way as to be able to demonstrate the new capabilities of this method. 
	
	\section{Preliminaries}
	In this paper vectors and matrices will be used. If a matrix is created from vectors, the vectors will be the columns of that matrix. The tensor concept will be introduced and its rotation will be described too. Only first and second rank tensors\footnote{It should be noted that despite the objections, publications interchangeably uses the terms "rank of tensor" and "order of tensor" \cite{sochi2016intro}.} will be used in this article \cite{sochi2016intro}. For description of the operations on tensors the matrix notation will be used, instead of dummy indexes \cite{JFN57}.The terms "coordinate system" and "the base" will be used interchangeably.
	
	\subsection{Used abbreviations and symbols}
	It is assumed that if a vector or matrix or any other object has the "prime" symbol ($'$), this object is described in a coordinate system (the base) other than the standard coordinate system. If there is no this symbol, the object is given in the standard coordinate system:

	\begin{table}[H]
	\centering
	\fontsize{10}{12}\selectfont{
		\begin{tabular}{l l c} 
		$a$ & $-$ & \parbox[t]{11.5 cm}{a vector representing a single point in the space of standardized primary random variables (one row in the input data table).}\\
		$A$ & $-$ & \parbox[t]{11.5 cm}{matrix of vectors representing standardized primary variables in a standard base (standard coordinate system) Note: The column in matrix $A$ can not be identified with a single point in the data space (previously denoted as $a$).}\\
		$A'$ & $-$ & \parbox[t]{11.5 cm}{matrix of vectors representing standardized primary variables in the base of eigenvectors.}\\
		$C$ & $-$ & \parbox[t]{11.5 cm}{correlation coefficient matrix.}\\
		$C'$ & $-$ & \parbox[t]{11.5 cm}{correlation coefficient matrix after diagonalization (the matrix of correlation coefficients in the base of eigenvectors).}\\
		$I$ & $-$ & \parbox[t]{11.5 cm}{a identity matrix containing all axes of the standard base.}\\
		$p$ & $-$ & \parbox[t]{11.5 cm}{a vector representing a single point in the principal components space (one row in the principal components data table).}\\
		$p_{ci}$ & $-$ & \parbox[t]{11.5 cm}{$i-$th principal component or $i-$th vector representing this principal component.}\\
		$P$ & $-$ & \parbox[t]{11.5 cm}{matrix of vectors representing principal components in the standard coordinate system (standard base).}\\
		$P'$ & $-$ & \parbox[t]{11.5 cm}{matrix of vectors representing principal components in the base of eigenvectors. Note: The column in matrix $P'$ can not be identified with a single point in the princiapal components data table (previously denoted as $p$).}\\
		$R$ & $-$ & \parbox[t]{11.5 cm}{rotation matrix described transition from standard coordinate system.}\\
		$u_i$ & $-$ & \parbox[t]{11.5 cm}{$i-$th vector representing the axis of the coordinate system after rotation ($i-$th eigenvector).}\\
		$U$ & $-$ & \parbox[t]{11.5 cm}{matrix containing directional vectors of coordinate system after rotation (matrix of eigenvectors).}\\
		$v$ & $-$ & \parbox[t]{11.5 cm}{vector in standard coordinate system (standard base).}\\
		$v'$ & $-$ & \parbox[t]{11.5 cm}{vector in the base after rotation (in rotated coordinate system).}\\
\end{tabular}}
\end{table}
	
	\subsection{The concept of tensor}
	For description of the tensor concept, Cartesian coordinate system is assumed. Quantities, which are not dependent on this coordinate system, are scalars. Scalar is described by one number. On the contrary, some other quantities are defined with respect to coordinate system. These quantities are tensors. Tensor is a quantity, which is described by a number of components, whose values are depending on the coordinate system. In tensor notation, tensor rank manifests by number of indices:

	\begin{itemize}
		\item Scalar is a tensor of rank zero and it has no index.
		\item Vector is a first rank tensor and has only one index. For the given coordinate system in $n-D$ space, vector is completely defined by their $n$ components. These elements are perpendicular projections of the vector to the respective axes of the system.
		\item Tensor of rank two has two indices. It is a quantity, which is defined by $n\times n$ numbers, which form a square matrix. As examples of the second rank tensor, a matrix of quadratic form can be used as well as a matrix of correlation coefficients. At this point it should be noted that second rank tensors are represented by square matrices, but not all square matrices are tensors \cite{sochi2016intro}.
	\end{itemize}
	Higher rank tensors can be also considered, but they are not the subject of this study.
		
	\subsubsection{Rotation of coordinate system}
		
\begin{figure}
	\centering
	\includegraphics[width=9cm]{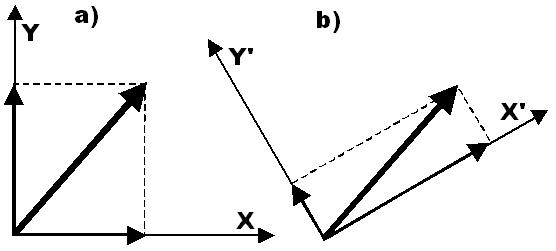}
	\caption{Rotation of Cartesian coordinate system. Components of vector: a) before coordinate system rotation, b) after coordinate system rotation}\label{fig1}
\end{figure}

\begin{table}
	\centering
	\caption{Direction cosines between axes before and after rotation}\label{tab1}
	\fontsize{10}{14}\selectfont{
		\begin{tabular}{c|c|c|c|c} \hline 
			\multicolumn{2}{c|}{\ } & \multicolumn{3}{|c}{Axes before rotation} \\ \cline{3-5}
			\multicolumn{2}{c|}{\ } & $X_1$ & \ldots & $X_n$ \\ \hline \hline
			Axes & $X_1'$ & $cos(X_1',X_1)$ & \ldots & $cos(X_1',X_n)$ \\ \cline{2-5}
			after & \vdots & \vdots & \vdots & \vdots \\ \cline{2-5}
			rotation & $X_n'$ & $cos(X_n',X_1)$ & \ldots & $cos(X_n',X_n)$ \\ \hline 
	\end{tabular}}
\end{table}

	As an example vector on a plane (first rank tensor in $2-D$) can be considered (see Fig. \ref{fig1}). This vector observed in different coordinate systems (before rotation and after rotation) is the same vector. In these different coordinate systems it has different components (projections on coordinate system axes). For known vector components in the coordinate system before rotation, the vector components in the coordinate system after rotation should be found.
	For solving problem of changes of tensor components, transformation of coordinate system has to be described. Transformation of tensor component can be described after them.
	
	Rotation of coordinate system without change of origin is considered. Axes before rotation are denoted as $X_1,X_2,...,X_n$. After rotation the same axes are denoted as $X_1',X_2',...,X_n'$. Table \ref{tab1} shows direction cosines between coordinate system axes before and after rotation.
	To find this table, the set of cosines between vectors should be calculated using formula:
	
	\begin{equation}\label{Eq1}
		cos(X_i',X_j)=\frac{X_i'\cdot X_j}{\left\|X_i'\right\|_2 \cdot \left\|X_j\right\|_2 }.
	\end{equation}
	Since standard base vectors as well as vectors of finale base have unit lengths, the denominator in the above expression is equal to one and the cosine is equal to the scalar product of the corresponding vectors.
	
	The content of Table \ref{tab1} forms a rotation matrix. Denoting $cos(X_i',X_j)$ as $r_{ij}$, rotation matrix is a square matrix of $n\times n$ dimension, which is denoted as $R$:
	\begin{equation}\label{Eq2}
		R {=}
		\begin{bmatrix}
		r_{11} & \ldots & r_{1n} \\ 
		\vdots & \ddots & \vdots \\ 
		r_{n1} & \ldots & r_{nn} \end{bmatrix}.
	\end{equation}
	Components of rotation matrix $R$ are mutually dependent. Multiplying transposition matrix $R^T$ by matrix $R$, identity matrix is achieved. It means matrix $R$ is an orthogonal matrix:
	\begin{equation}\label{Eq3}
		R^T=R^{-1}.
	\end{equation}
	
	Let the original base is a standard coordinate system represented by the $n$ vectors $b_1,b_2,\ldots,b_n$ such that all components of $b_i$ are all zero, apart from the unit element with the index $i$. As the columns of the first matrix the successive standard base vectors will be inserted. These columns will form the identity matrix :
	\begin{equation}\label{Eq4}
		I {=}
		\begin{bmatrix}
		1 & \ldots & 0 \\ 
		\vdots & \ddots & \vdots \\ 
		0 & \ldots & 1 \end{bmatrix}.
	\end{equation}
	On the other hand, the final base should be a set of orthogonal vectors $u_1,u_2,\ldots,u_n$ such that $u_i=[u_{1i},\ldots , u_{ni}]^T$ and $\left\| u_i\right\| _2=1$. Let successive vectors of the new base become the successive columns of the second matrix:
	
	\begin{equation}\label{Eq5}
		U {=}
		\begin{bmatrix}
			u_{11} & \ldots & u_{1n} \\ 
			\vdots & \ddots & \vdots \\ 
			u_{n1} & \ldots & u_{nn} 
		\end{bmatrix}.
	\end{equation}
	In the algorithm of finding the product of two matrices, the element with indices $i$ and $j$ in the resulting matrix is the scalar product of the $i-$th row in the first matrix and the $j-$th column in the second matrix. Because of the unit lengths of column vectors in matrices (\ref{Eq4}) and (\ref{Eq5}), the problem of finding the $R$ rotation matrix is reduced to finding the product of two matrices:
	
	\begin{equation}\label{Eq6}
		R=U^T I.
	\end{equation}
	Hence, the transition (rotation) matrix from the primary to the final base is a matrix whose rows are the directional vectors of the final base:
	\begin{equation}\label{Eq7}
		R=U^T {=}
		\begin{bmatrix}
			u_{11} & \ldots & u_{n1} \\ 
			\vdots & \ddots & \vdots \\ 
			u_{1n} & \ldots & u_{nn} 
		\end{bmatrix}.
	\end{equation}
	
	\subsubsection{Transformation of tensor component}
	
	\begin{table}
		\centering
		\caption{Transformation of tensors in matrix notation}\label{tab2}
		\fontsize{10}{14}\selectfont{
		\begin{tabular}{c|c|c|c} \hline 
			Rank & New components & Old components & \multirow{2}{*}{Note} \\ 
			of	tensor & expressed by old & expressed by new & \\ \hline \hline
			0	& $\varphi '=\varphi$ & $\varphi =\varphi '$ & $\varphi$, $\varphi '$ $-$ scalar \\ \hline 
			{1} & {$v'=R v$} & {$v=R^T v'$} & {$v$, $v'$ $-$ vector} \\ \hline 
			\multirow{2}{*}{2} & \multirow{2}{*}{$C'=R C R^T$} & \multirow{2}{*}{$C=R^T C' R$} & $C$, $C'$ $-$ second rank tensor \\ 
			& & & expressed as a square matrix\\ \hline
		\end{tabular}}
	\end{table}
	
	Vector $V=[v_1,v_2,\ldots ,v_n ]^T$ is considered in some Cartesian coordinate system. If vector components before rotation are given and transformation $R$ is known, then it is possible to find the set of vector components in the new coordinate system. If coordinate system is rotated with the use of matrix (\ref{Eq2}), then vector $v$ will be observed as a vector $v'$ with new components $v'=[v_1',v_2',\ldots,v_n' ]^T$, in the new coordinate system. The change of vector components is described by the formula \cite{JFN57}:
	\begin{equation}\label{Eq8}
		v'=R v.
	\end{equation}
	To return from the new coordinate system to the old one, both sides of equation (\ref{Eq8}) should be multiplied on the left by the inverse (transposition) of matrix $R$:
	\begin{equation}\label{Eq9}
		v=R^T v'.
	\end{equation}
	
	Since components of vector are dependent on coordinate system, in the same way components of higher rank tensors are also dependent on coordinate system. If coordinate system is rotated then tensor components are changed with this rotation. If tensor components before rotation are given and transformation of coordinate system (\ref{Eq2}) is known, then it is possible to find the set of tensor components in the new coordinate system. In Table \ref{tab2}, formulas for transformation of tensor components up to rank two are presented \cite{JFN57}.
	
	\subsection{Geometric interpretation of determination coefficient}\label{SubSection: Geometric interpretation of determination}
			
\begin{figure}
	\centering
	\includegraphics[width=8cm]{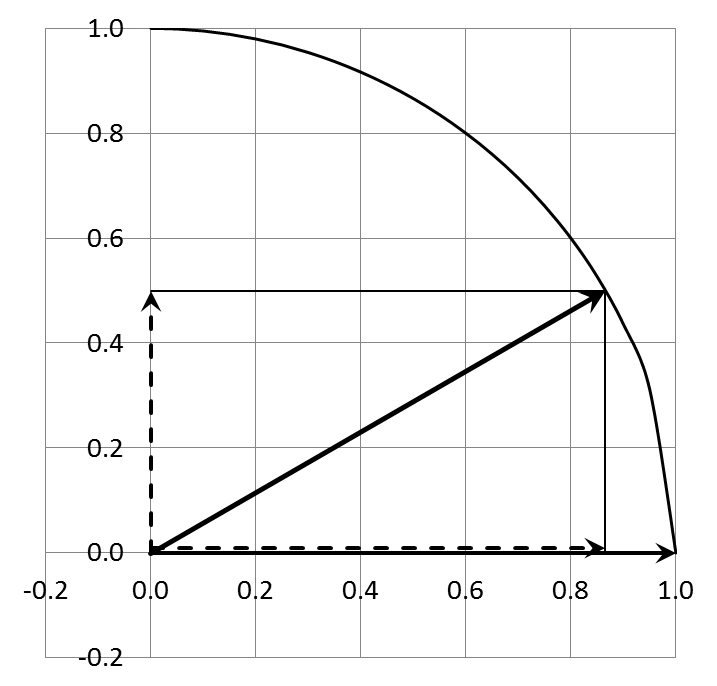}
	\caption{Geometric interpretation of determination coefficient}\label{fig2}
\end{figure}
	
	A measure of the relation between two random variables $X$ and $Y$ is the correlation coefficient. Denoting the random components of both variables as $x=X-\overline{X}$ and $y=Y-\overline{Y}$, the correlation coefficient can be expressed as the ratio of the scalar product of these vectors to the product of their lengths:

	\begin{equation}\label{Eq10}
	\rho_{X,Y}=\frac{\sum_{i=1}^{n}x_i y_i}{\sqrt{\sum_{i=1}^{n}x_i^2} \sqrt{\sum_{i=1}^{n}y_i^2}}=\frac{x\cdot y}{\left\| x\right\|_2 \cdot \left\| y\right\|_2 }=cos(x,y).
	\end{equation}
	This coefficient expresses the cosine of the angle between the random components of both variables \cite{gniazd2013}.
	Its square is called the coefficient of determination \cite{taylor1990}. In order to present the geometric interpretation of the determination coefficient, the following facts should be noted:
	\begin{itemize}
		\item The standard deviation of the random variable is equal to the root of its variance \cite{Papoulis1965}. This means that a random variable with a unit variance also has a unit standard deviation. This fact applies to the standardized random variables.
		\item If the random variable is a sum of independent random variables, then its variance is equal to the sum of the variances of the added variables \cite{Papoulis1965}.
		\item Standardization of the variable does not affect the value of the correlation coefficient.
	\end{itemize}
	Two standardized random variables with known correlation coefficients are considered. Symbolically, they can be represented as vectors of unit lengths placed on a circle with a radius of one (see Fig. \ref{fig2}). The vector lying along the horizontal axis represents the explanatory variable. The vector set at an angle represents the variable to be explained. The angle between these vectors is chosen so that its cosine is equal to the correlation coefficient between these two variables. Lengths of the vectors represent the standard deviations of (standardized) random variables. Squares of lengths of vectors represent the variances of these variables. Since vectors have unit lengths, the cosine (the correlation coefficient) is represented by the length of the projection of the explained vector per the explanatory vector.

	There is a question, what part of the variance of the explained variable is represented by the explanatory variable. 
	The ratio of the square of the length of the projection of explained vector per the explanatory vector (on horizontal axis) to the square of the length of the explanatory vector is the coefficient of determination. This coefficient indicates what part of the variance of the explained variable is explained by the variance of the explanatory variable \cite{taylor1990}.
	It is a number from the range $\left\langle 0,1\right\rangle $, which can also be expressed in percentages. 
	Because of the symmetry of the determination coefficient, the presented reasoning can be reversed by changing the explained variable with the explanatory variable.
	From here it can also be said that the coefficient of determination describes the level of a common variance of two correlated random variables.
	
	From Pythagorean theorem, the square of the length of the explained vector can be expressed as a sum of the squares of the lengths of its projections per orthogonal axes (Figure \ref{fig2}). One axis is consistent with the direction of the explanatory vector and the second axis is in the orthogonal direction. Orthogonal projections indicated by dashed lines represent independent random variables. 
	Just as the random variable can be represented as the sum of independent random variables, its variance can be represented as the sum of the variances of these (summed) variables.
	Just as the vector representing an explained variable can be represented as the sum of its orthogonal components, the square of the length of that vector is the sum of the squares of lengths of those components.
	Like the ratio of the projection of the explained vector to the explanatory vector is the cosine of the angle between these two vectors, the components of explained vector are the correlation coefficients between the explained variable and the independent explanatory variables represented by the orthogonal vectors.
	
	The above conclusion can be generalized to a multidimensional space. If there is an explained variable and there is a set of explanatory variables, then there is an orthogonal base where the vector representation of the explained variable will have components equal to the correlation coefficients between the explained variable and successive explanatory variables.
	
	\section{Principal Component Analysis}
	The matrix of correlation coefficients is used for finding the principal components. For this correlation matrix, the eigenproblem is solved. Several of the largest eigenvalues can to explain the most of the variation of analysed random variables. The original set of $n$ mutually correlated random variables can be represented by a smaller set of independent hypothetical variables. This new set contains less variables than the original dataset. That means that there is a space in which several hypotetical variables adequately explain the random behavior of the analyzed primary dataset \cite{hand2001}\cite{Larose2006DMmethods}.
	\subsection{Data for analysis}
	
	\begin{table}
		\centering
		\caption{The data for analysis (in centimeters)}\label{tab3}
		\fontsize{10}{14}\selectfont{
			\begin{tabular}{c|c|c|c|c} \hline 
				Sepal Length & Sepal Width & Petal Length & Petal Width & Class \\ \hline \hline
				$5.1$ & $3.5$ & $1.4$ & $0.2$ & Iris-setosa \\ \hline
				$4.9$ & $3$ & $1.4$ & $0.2$ & $\ldots$ \\ \hline
				$4.7$ & $3.2$ & $1.3$ & $0.2$ & $\ldots$ \\ \hline
				$\vdots$ & $\vdots$ & $\vdots$ & $\vdots$ & $\vdots$ \\ \hline
				$6.4$ & $3.2$ & $4.5$ & $1.5$ & Iris-versicolor \\ \hline
				$6.9$ & $3.1$ & $4.9$ & $1.5$ & $\ldots$ \\ \hline
				$5.5$ & $2.3$ & $4$ & $1.3$ & $\ldots$ \\ \hline
				$\vdots$ & $\vdots$ & $\vdots$ & $\vdots$ & $\vdots$ \\ \hline
				$6.3$ & $3.3$ & $6$ & $2.5$ & Iris-virginica \\ \hline
				$5.8$ & $2.7$ & $5.1$ & $1.9$ & $\ldots$ \\ \hline
				$7.1$ & $3$ & $5.9$ & $2.1$ & $\ldots$ \\ \hline
				$\vdots$ & $\vdots$ & $\vdots$ & $\vdots$ & $\vdots$ \\ \hline \hline
				$5.845$ & $3.121$ & $3.770$ & $1.199$ & Average \\ \hline
				$0.833$ & $0.480$ & $1.773$ & $0.763$ & Standard deviation \\ \hline
				$0.693$ & $0.230$ & $3.143$ & $0.582$ & Variance \\ \hline
		\end{tabular}}
	\end{table}
	
	\begin{table}
		\centering
		\caption{The matrix of correlation coefficient (cosines)}\label{tab4}
		\fontsize{10}{14}\selectfont{
			\begin{tabular}{c|c|c|c|c} \hline 
				& Sepal Length & Sepal Width & Petal Length & Petal Width \\ \hline \hline
				Sepal Length & $1$ & $-0.063$ & $0.866$ & $0.816$ \\ \hline
				Sepal Width & $-0.063$ & $1$ & $-0.321$ & $-0.300$ \\ \hline
				Petal Length & $0.866$ & $-0.321$ & $1$ & $0.959$ \\ \hline
				Petal Width & $0.816$ & $-0.300$ & $0.959$ & $1$ \\ \hline
				
		\end{tabular}}
	\end{table}
	
	\begin{table}
		\centering
		\caption{The matrix of significance levels}\label{tab5}
		\fontsize{10}{14}\selectfont{
			\begin{tabular}{c|c|c|c|c} \hline 
				& Sepal Length & Sepal Width & Petal Length & Petal Width \\ \hline \hline
				Sepal Length & $0$ & $0.446$ & $0.000$ & $0.000$ \\ \hline
				Sepal Width & $0.446$ & $0$ & $0.000$ & $0.000$ \\ \hline
				Petal Length & $0.000$ & $0.000$ & $0$ & $0.000$ \\ \hline
				Petal Width & $0.000$ & $0.000$ & $0.000$ & $0$ \\ \hline
				
		\end{tabular}}
	\end{table}
	
	\begin{table}
		\centering
		\caption{The matrix of angles given in degrees}\label{tab6}
		\fontsize{10}{14}\selectfont{
			\begin{tabular}{c|c|c|c|c} \hline 
				& Sepal Length & Sepal Width & Petal Length & Petal Width \\ \hline \hline
				Sepal Length & $0$ & $93.59$ & $30.05$ & $35.29$ \\ \hline
				Sepal Width & $93.59$ & $0$ & $108.74$ & $107.47$ \\ \hline
				Petal Length & $30.05$ & $108.74$ & $0$ & $16.44$ \\ \hline
				Petal Width & $35.29$ & $107.47$ & $16.44$ & $0$ \\ \hline
		\end{tabular}}
	\end{table}
	
	\begin{table}
		\centering
		\caption{The matrix of determination coefficients}\label{tab7}
		\fontsize{10}{14}\selectfont{
			\begin{tabular}{c|c|c|c|c} \hline 
				& Sepal Length & Sepal Width & Petal Length & Petal Width \\ \hline \hline
				Sepal Length & $100.00\%$ & $0.39\%$ & $74.93\%$ & $66.63\%$ \\ \hline
				Sepal Width & $0.39\%$ & $100.00\%$ & $10.32\%$ & $9.01\%$ \\ \hline
				Petal Length & $74.93\%$ & $10.32\%$ & $100.00\%$ & $91.99\%$ \\ \hline
				Petal Width & $66.63\%$ & $9.01\%$ & $91.99\%$ & $100.00\%$ \\ \hline
		\end{tabular}}
	\end{table}
	
	As a data for the analysis the flower Iris data proposed by Sir Ronald Fisher in 1930 will be used \cite{fisher1936}. On the one hand, these data are not too difficult to analyze. On the other hand, this data are enough to show many aspects of the principal component analysis. 
	This data table contains four columns with correlated numeric data and a fifth column with nominal data. Only numeric columns will be used in this article. Table \ref{tab3} lists several records of these data. At the bottom of the columns with numbers the mean, standard deviation, and the variance for each column are presented. 		
	
	For considered data the matrix of correlation coefficients was calculated (Table \ref{tab4}). The significance level \cite{gniazd2013} for each correlation coefficient was also examined (Table \ref{tab5}). Except in one case, all correlations are significant on the level less than 0.001.
	Table \ref{tab6} shows the angles in degrees calculated on the basis of the cosine (correlation coefficients) between the components of the considered random variables \cite{gniazd2013}. 
	Angles close to the right angle confirms the strong independence of random variables. 
	Angles of close to zero or 180 degrees confirm the strong dependence of random variables. 
	The table shows that the second variable (Sepal Width) is almost orthogonal to the other variables.
	On the basis of the matrix of correlation coefficients were also found corresponding coefficients of determination \cite{taylor1990}.
	Table \ref{tab7} shows the results given as a percentage.
	The coefficients of determination between the second and the other variables do not exceed $11\%$. 
	On the other hand, the determination coefficients between the first, third and fourth variables are not less than $66\%$
	
	\subsection{Reconstruction of principal components}\label{SubSection: subset of PC}
		
	\begin{figure}
		\centering
		\includegraphics[width=9cm]{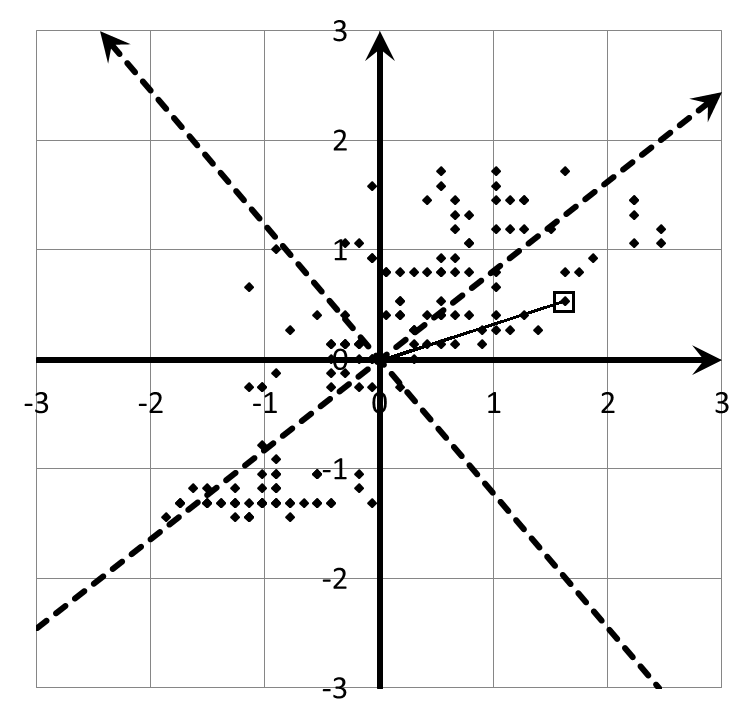}
		\caption{Rotation of the coordinate system. The point surrounded by the square will have different coordinates in the standard coordinate system (the axes marked with solid lines) and in the system of eigrnvectors (the axes marked with dashed lines)}\label{fig3}
	\end{figure}
	
	\begin{table}
		\centering
		\caption{Eigenvectors in columns}\label{tab8}
		\fontsize{10}{14}\selectfont{
			\begin{tabular}{c|c|c|c} \hline 
				$0.534$ & $0.317$ & $0.757$ & $0.203$ \\ \hline
				$-0.213$ & $0.948$ & $-0.229$ & $-0.066$ \\ \hline
				$0.584$ & $0.026$ & $-0.212$ & $-0.783$ \\ \hline
				$0.573$ & $0.030$ & $-0.574$ & $0.584$ \\ \hline
		\end{tabular}}
	\end{table}
	
	\begin{table}
		\centering
		\caption{Several objects in the principal components space}\label{tab9}
		\fontsize{10}{14}\selectfont{
			\begin{tabular}{c|c|c|c|c} \hline 
				No. & $p_{c1}$ & $p_{c2}$ & $p_{c3}$ & $p_{c4}$ \\ \hline \hline
				$1$ & $-2.184$ & $0.393$ & $-0.176$ & $0.048$ \\ \hline
				$2$ & $-2.091$ & $-0.674$ & $-0.233$ & $0.068$ \\ \hline
				$3$ & $-2.341$ & $-0.356$ & $0.033$ & $0.036$ \\ \hline
				$\vdots$ & $\vdots$ & $\vdots$ & $\vdots$ & $\vdots$ \\ \hline \hline
				Average & $0.000$ & $0.000$ & $0.000$ & $0.000$ \\ \hline
				Standard deviation & $1.694$ & $0.984$ & $0.398$ & $0.181$ \\ \hline
				Variance & $2.868$ & $0.967$ & $0.159$ & $0.033$ \\ \hline
		\end{tabular}}
	\end{table}
	
	Based on the matrix of correlation coefficients, the principal components were analyzed. 
	As a result of the solution of the problem, the correlation coefficient matrix was diagonalized.
	The following eigenvalues were obtained: $2.849$, $0.961$, $0.158$ and $0.033$. 
	The corresponding eigenvectors with length reduced to the unity are shown in the columns in Table \ref{tab8}. 
	The order of eigenvectors corresponds to eigenvalues sorted from largest to smallest.
	Orthogonal eigenvectors represent the new base in which the primary random variables will be represented. 
	Transposed matrix of eigenvectors creates an orthogonal rotation matrix (\ref{Eq7}).
	This matrix will be used to find mutually independent principal components.
	
	Each row in the table of standardized primary variables (Table \ref{tab3}) is a vector representing one point in the primary variables space. 
	Its projection to the eigenvectors (new axes of the system) will give a point in the principal components space. 
	Denoting by $a^T$ the row from the standardized primary data (Table \ref{tab3}) and by $p^T$ the unknown row in the principal components space, and using the matrix $R$, we can accomplish this in the following way:
	\begin{equation}\label{Eq11}
	p^T=R a^T.
	\end{equation}
	Transformation (\ref{Eq11}) is equivalent to the rotation of the standard coordinate system (standard base) to the coordinate system defined by the eigenvectors. 
	Before the rotation the row vector is known in the standard base as the vector $a^T$. 
	After making the transformation (\ref{Eq11}), its description is found in the base of eigenvectors.
	For example, in Figure \ref{fig3}, the point (surrounded by a square) with coordinates $(1,632,0.528)$ in the standard coordinate system, after rotation of this system becomes a point with coordinates $(1,899, -0.243)$.
	Repeating procedure (\ref{Eq11}) for all points in the space of standardized primary variables gives all points in the space of principal components. 
	Table \ref{tab9} shows several objects in the spaces of principal components.
	At the bottom of this table are given mean values, standard deviations and variances for each column. 
	It is noted that the variances of each variable $p_{c1}, \ldots, p_{c4}$ are (with an accuracy of calculation errors) equal to the previously obtained eigenvalues.
	
	\subsection{Choosing a subset of principal components}
	
	\begin{figure}
		\centering
		\includegraphics[width=8cm]{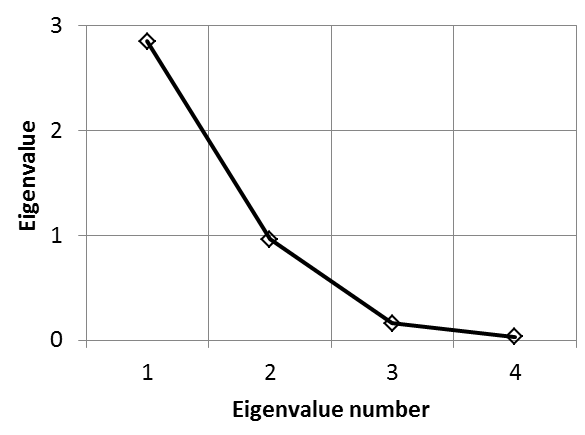}
		\caption{The scree plot}\label{fig4}
	\end{figure}
	
	\begin{table}
		\centering
		\caption{The percentage of variance explained by the successive principal components (in brief: PC)}\label{tab10}
		\fontsize{10}{14}\selectfont{
			\begin{tabular}{c|c|c|c|c} \hline 
				\multirow{2}{*}{No.} & \multirow{2}{*}{Eigenvalue} & Cumulative & Percentage of variance & Cumulative \\ 
				& & eigenvalues & explained by each PC & percentage of variance \\ \hline \hline
				$1$ & $2.849$ & $2.849$ & $71.22\%$ & $71.22\%$ \\ \hline
				$2$ & $0.961$ & $3.810$ & $24.02\%$ & $95.24\%$ \\ \hline
				$3$ & $0.158$ & $3.967$ & $3.94\%$ & $99.19\%$ \\ \hline
				$4$ & $0.033$ & $4.000$ & $0.81\%$ & $100.00\%$ \\ \hline
				
		\end{tabular}}
	\end{table}
	
	An important issue is the choice of the amount of the extracted principal components.
	Various criteria are used for this purpose \cite{Larose2006DMmethods}:
	\begin{itemize}
		\item Percentage criterion of the part of the variance which is explained by the principal components.
		It is assumed that there are so many principal components that the sum of the eigenvalues associated with the successive principal component is not less than the established threshold relative to the trace of the correlation coefficient matrix.
		\item Criterion of a scree plot (see Fig. \ref{fig4}). 
		The plot shows the successive eigenvalues, from largest to smallest. 
		Its shape resembles a scree. 
		It is as many principal components as eigenvalues located on the slope of the scree.
		\item Criterion of eigenvalue. 
		The number of principal components is equal to the number of eigenvalues of not less than one.
	\end{itemize}
	Because each of the above criteria may suggest another number of components, the final decision about their numbers is taken by the human.
	Table \ref{tab10} shows eigenvalues, cumulative eigenvalues, the percentage of variance explained by the principal components, and the cumulative percentage of the variance.
	This table shows that two principal components carry over $95\%$ of the information contained in primary variables. 
	Also, the scree plot shows that there are two values on its slope, while the other two are off the slope.
	According to the eigenvalue criterion, the second eigenvalue is slightly less than unity.
	By analyzing all the criteria for selecting the principal components presented above, it can be concluded that two principal components can be selected for representation of statistical behavior of a set of analyzed variables. 
	Such a quantity sufficiently explains at least $95\%$ of their variance.
		
	\section{Expandability of PCA capabilities}
	
	\begin{table}
		\centering
		\caption{Correlation coefficients between primary variables and principal components}\label{tab11}
		\fontsize{10}{14}\selectfont{
			\begin{tabular}{c||c|c|c|c} \hline 
				& Sepal Length & Sepal Width & Petal Length & Petal Width \\ \hline \hline
				$p_{c1}$ & $0.901$ & $-0.359$ & $0.986$ & $0.968$ \\ \hline
				$p_{c2}$ & $0.311$ & $0.929$ & $0.025$ & $0.030$ \\ \hline
				$p_{c3}$ & $-0.301$ & $0.091$ & $0.084$ & $0.228$ \\ \hline
				$p_{c4}$ & $0.037$ & $-0.012$ & $-0.141$ & $0.105$ \\ \hline
		\end{tabular}}
	\end{table}
	
	\begin{table}
		\centering
		\caption{Determination coefficients between primary variables and principal components}\label{tab12}
		\fontsize{10}{14}\selectfont{
			\begin{tabular}{c||c|c|c|c||c} \hline 
				&Sepal Length&Sepal Width&Petal Length&Petal Width & $\Sigma $ \\ \hline \hline
				$p_{c1} $ & $0.812$ & $0.129$ & $0.972$ & $0.936$ & $2.849$ \\ \hline
				$p_{c2} $ & $0.097$ & $0.863$ & $0.001$ & $0.001$ & $0.961$ \\ \hline
				$p_{c3} $ & $0.090$ & $0.008$ & $0.007$ & $0.052$ & $0.158 $\\ \hline
				$p_{c4} $ & $0.001$ & $0.000$ & $0.020$ & $0.011$ & $0.033 $ \\ \hline \hline
				$\Sigma$ & $1.000$ & $1.000$ & $1.000$ & $1.000$ & $4.000$ \\ \hline	
		\end{tabular}}
	\end{table}
	
	Formula (\ref{Eq11}) made it possible to find the principal components based on the standardized primary variables.
	In order to interpret the obtained results, correlation coefficients between primary variables and principal components were calculated (Table \ref{tab11}). 
	Based on the correlation, appropriate coefficients of determination were calculated. 
	For the given coefficients of determination the sums of the elements in columns and rows were calculated too (Table \ref{tab12}).

	\subsection{New interpretation of PCA in virtual vector space}
	The columns in the Tables \ref{tab11} and \ref{tab12} refer to standardized primary variables. On the other hand, rows in Tables \ref{tab11} and \ref{tab12} refer to the principal components. The consistent mutual analysis of both tables allows to find a new geometric interpretation of the principal component method.

	\subsubsection{Standardized primary variables - analysis of columns}
	Elements in the columns of Table \ref{tab12} sum to the unity, which is the variance of subsequent standardized primary variables.
	This means that the standardized primary variable is the sum of independent variables, and its variance is equal to the sum of variances of those variables (see Sec. \ref{SubSection: Geometric interpretation of determination}).
	It can be said that standardized primary variables divide their variance between independent principal components.
	For example, a standardized variable Sepal Length divides its variance between four mutually independent principal components $p_{c1}$, $p_{c2}$, $p_{c3}$ as well as $p_{c4}$.
	About $81\%$ of the variance of the variable Sepal Length is a part of the principal component $p_{c1}$, almost $10\%$ is a part of $p_{c2}$, about $9\%$ is passed to the principal component $p_{c3}$, and the remainder of variance (less than $1\%$) is a part of $p_{c4}$.
	
	The following columns of Table \ref{tab11} are the correlation coefficients between successive primary variables, and independent principal components.
	Figure \ref{fig2} shows that the correlation coefficient can be interpreted as a projection of a unit length vector per axis of the coordinate system. 
	This projection is one of the vector components. 
	Consequently, the correlation coefficients in a given column can be treated as components of a vector. 
	If the correlation coefficients in a given column of Table \ref{tab11} are taken as vector components, then by the Pythagorean theorem the sum of the squares of the vector projections lengths per perpendicular axes equals the square of the length of this vector.
	
	Principal components are associated with eigenvectors. 
	The components of subsequent column vectors in Table \ref{tab11} are projections for orthogonal eigenvectors. 
	This means that primary variables are represented as vectors described in the base of eigenvectors. These vectors are columns in the matrix $A'$:
	
	\begin {equation}\label {Eq12} 
	A'=
	\begin {pmatrix} 
	0.901&-0.359&0.986&0.968\\
	0.311&0.929&0.025&0.030\\
	-0.301&0.091&0.084&0.228\\
	0.037&-0.012&-0.141&0.105
	\end {pmatrix} 
	\end {equation} 
	
	\subsubsection{Principal components - analysis of rows}
	The elements in the rows of Table \ref{tab12} sums to eigenvalues that are the variances of the principal components. 
	The variance of the given principal component consists of mutually independent (orthogonal) parts of the variance of the standardized primary variables. 
	This means that this principal component consists of the sum of independent random variables, and its variance is equal to the sum of the variances of those variables.
	And so, to the variance of the principal component $p_{c1}$ the variables Sepal Length, Petal Length as well as Petal Width give the most.
	To the variance of the principal component $p_{c2}$ contributes most variable Sepal Width.
	The variances of the principal components $p_{c3}$ and $p_{c4}$ are much less than the variances of the principal components $p_{c1}$ and $p_{c2}$.
	
	In the subsequent rows of Table \ref{tab11} there are coefficients of correlation between successive principal components and primary variables. 
	By virtue of Pythagorean theorem, the sum of the squares of these components is equal to the square of the length of the vector (Table \ref{tab12}). 
	These are equal to the eigenvalues, so they are equal to variances of consecutive principal components.
	
	Primary variables are associated with a standard base. 
	Components of consecutive row vectors are projections on the orthogonal axes of the standard base. 
	This means that the principal components are represented as vectors in the standard base. 
	Rows in Table 11 are columns in the matrix P:
	
	\begin {equation}\label {Eq13} 
	P=
	\begin {pmatrix} 
	0.901&0.311&-0.301&0.037\\
	-0.359&0.929&0.091&-0.012\\
	0.986&0.025&0.084&-0.141\\
	0.968&0.030&0.228&0.105
	\end {pmatrix} 
	\end {equation} 
	\subsubsection{Vector representation of both standardized primary variables and principal components}
	As a consequence of the analysis, it was shown that both the standardized primary variables and the principal components can be seen as a vectors described in some bases. 
	The columns in Table \ref{tab11} are vectors representing standardized primary variables in the eigenvectors base. 
	These columns form the matrix $A'$. 
	On the other hand, the rows in Table \ref{tab11} are vectors representing the principal components in the standard base. 
	The transposition of the rows of Table \ref{tab11} forms the matrix $P$. 
	Between the representation of the standardized primary variables (matrix $A'$ in the base of eigenvectors) and the representation of the principal components (Matrix $P$ in standard base) there is a relation:
	\begin{equation}\label{Eq14}
	A'=P^T.
	\end{equation}
	For vectors representing standardized primary variables, the natural base is the standard base. 
	For vectors representing principal components, the natural base is the base of the eigenvectors. 
	The question is: What is their character in their natural bases.
	
	\paragraph{Standardised primary variables in the natural base}
	It is known representation $A'$ of the standardized primary variables (\ref{Eq12}) in the base of eigenvectors. 
	On the other hand, it has been shown that the eigenvectors are the rows of a rotation matrix $R$. This matrix (or its transposition) describes the transition between the standard base and the eigenvectors base. 
	The equation (\ref{Eq9}) was used to find description of matrix $A$ in the standard base:
	\begin{equation}\label{Eq15}
	A=R^T A'.
	\end{equation}
	As a result of this transformation, the symmetric matrix $A$ was obtained:
	\begin {equation}\label {Eq16} 
	A=
	\begin {pmatrix} 
	0.815&0.032&0.442&0.375\\
	0.032&0.978&-0.157&-0.132\\
	0.442&-0.157&0.705&0.532\\
	0.375&-0.132&0.532&0.748
	\end {pmatrix} 
	\end {equation} 
	
	\paragraph{Principal components in the natural base}
	For vectors representing principal components, the natural base is the base of eigenvectors. 
	Vectors describing these principal components in the standard base are known. 
	These vectors forms the matrix $P$.
	To find the representation of the principal components in the eigenvectors base, formula (\ref{Eq8}) was used:
	\begin{equation}\label{Eq17}
	P'=R P.
	\end{equation}
	As a result, the diagonal matrix $P'$ was obtained:
	\begin {equation}\label {Eq18} 
	P'=
	\begin {pmatrix} 
	1.688&0.000&0.000&0.000\\
	0.000&0.980&0.000&0.000\\
	0.000&0.000&0.397&0.000\\
	0.000&0.000&0.000&0.180
	\end {pmatrix} 
	\end {equation} 
	When the matrix (\ref{Eq18}) is raised to the square, on its diagonal appear the variances of the principal components. This means that the squares of the lengths of the vectors representing the principal components are equal to the variances of these principal components.
	Obtained result is consistent with intuition:
	\begin{itemize}
		\item The principal components are represented by orthogonal vectors conforming to the directions of eigenvectors.
		\item The values on the diagonal are equal (with precision of numerical errors) to the standard deviations of the principal components.
	\end{itemize}
	
	\subsubsection{Ability to reconstruction of correlation matrix}
	It is assumed that in matrix $A$ containing vectors representing standardized primary variables, complete information is provided about the statistical properties of these variables. 
	Therefore, using this matrix, correlation coefficients between variables can be recreated. 
	Correlation coefficients have the interpretation of cosines of angles. 
	If the coordinates of individual vectors in a certain base are given, the cosines between the vectors can be found. 
	Since vectors representing standardized primary variables have unit lengths, the corresponding cosines are equal to the scalar product of the corresponding vectors. 
	Therefore, using the vector representation of the primary variables $A$, the matrix $C$ containing the correlation coefficients of the primary variables can be calculated by performing the following matrix multiplication:
	\begin{equation}\label{Eq19}
	C=A^T A.
	\end{equation}
	This formula is analogous to the formula (\ref{Eq10}) for the correlation coefficient of two random variables with a precision of skipped product of the lengths of vectors.
	
	It should be noted that the operation (\ref{Eq19}) can be performed both for vectors represented in the standard base (columns of the matrix (\ref{Eq16})) as well as in the base formed by the eigenvectors (columns of the matrix (\ref{Eq12})).
	
	\subsection{The level of explanation of primary variables}
	
	\begin{table}
		\centering
		\caption{Level of reconstruction of primary variables}\label{tab13}
		\fontsize{10}{14}\selectfont{
			\begin{tabular}{c||c|c|c|c||c} \hline 
				&Sepal Length&Sepal Width&Petal Length&Petal Width & Average in row \\ \hline \hline
				$p_{c1} $ & $81.17\%$ & $12.88\%$ & $97.23\%$ & $93.61\%$ & $71.22\%$ \\ \hline
				$p_{c2} $ & $9.65\%$ & $86.28\%$ & $0.06\%$ & $0.09\%$ & $24.02\%$ \\ \hline \hline
				$\Sigma$ & $90.82\%$ & $99.16\%$ & $97.29\%$ & $93.70\%$ & $95.24\%$ \\ \hline
		\end{tabular}}
	\end{table}
	
	Proposed in section \ref{SubSection: subset of PC} two principal components will adequately explain at least $95\%$ of the variance of the set of analyzed variables.
	The level of explanation of primary variables by the selected two principal components applies to all variables, not to individual variables. 
	On the other hand, it can be expected that the variance of each primary variable will be explained by selected principal components to varying degrees. 
	Since Table \ref{tab10} refers to all variables rather than to individual variables, additional analysis using determination coefficients is therefore needed.
	By reducing the determination coefficient table (Table \ref{tab12}) to two rows and giving results in percent (Table \ref{tab13}), you can estimate the level of reproduction of the variance of primary variables by the selected principal components. 
	Thus, the first primary variable will be retained in more than $90$ percent, the second variable in over $99$ percent, the third variable in over $97$ percent, and the fourth variable in over $93$ percent. 
	For individual primary variables, the loss of information is less than $10\%$, less than $1\%$, less than $3\%$, and less than $7\%$, respectively.
	It can also be seen that average values in the rows (last column in Table \ref{tab13}) are identical to the percentage values in the columns in Table \ref{tab10}. 
	The first two values agree with the percent variance explained by the principal components $p_{c1}$ and $p_{c2}$, and the third value agrees with the corresponding cumulative value (the cumulative percentage of the variance for two principal components).
	This means that the result of $95\%$ obtained in Table \ref{tab10} is only the average result for all variables.
	The use of this criterion refers to the mean level of explanation of the variables.
	
	In practice it may happen that some of the primary variables will be underrepresented.
	An additional criterion for selecting the appropriate number of principal components should therefore be used.
	This criterion should take into account the level of reconstruction of individual primary variables, as in the last row in Table \ref{tab13}.
	The number of principal components should be such that the representation level of each variable is at least satisfactory.

	\subsection{The concept of clustering of primary variables based on common variance}
		
	\begin{figure}
		\centering
		\includegraphics[width=7cm]{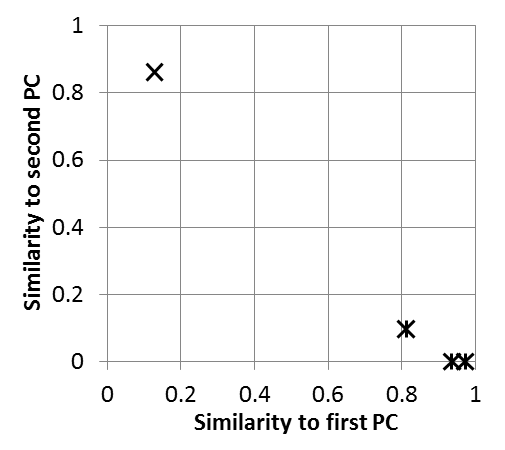}
		\caption{The similarity of the primary variables and selected principal components. Two different clusters are labeled with different markers.}\label{fig5}
	\end{figure}
	
	\begin{table}
		\centering
		\caption{The similarity of the primary variables and selected principal components measured by the coefficient of determination
		}\label{tab14}
		\fontsize{10}{14}\selectfont{
			\begin{tabular}{c||c|c|c|c} \hline 
				&Sepal Length&Sepal Width&Petal Length&Petal Width \\ \hline \hline
				Similarity to $p_{c1}$ & $0.812$ & $0.129$ & $0.972$ & $0.936$\\ \hline
				Similarity to $p_{c2}$ & $0.097$ & $0.863$ & $0.001$ & $0.001$\\ \hline
		\end{tabular}}
	\end{table}
		
	\begin{table}
		\centering
		\caption{Operations on tensors}\label{tab15}
		\fontsize{10}{14}\selectfont{
			\begin{tabular}{c|c|c} \hline 
				Transition &Back &\multirow{3}{*}{Note} \\ 
				to the new& to the old&\\ 
				coordinate system&coordinate system & \\ \hline\hline
				\multirow{3}{*}{$A'=RA$} & \multirow{3}{*}{$A=R^T A'$} & Standardized primary variables \\
				& & are represented as vectors \\
				& & (columns in matrices $A$ and $A'$)\\ \hline 
				\multirow{3}{*}{$P'=RP$} & \multirow{3}{*}{$P=R^T P'$} & Principal components \\
				& & are represented as vectors \\ 
				& & (columns in matrices $P$ and $P'$)\\ \hline
				\multirow{2}{*}{$C'=RCR^T $} &\multirow{2}{*}{$C=R^T C' R$} & The correlation coefficient matrix \\
				& & ($C$, $C'$) is a tensor of rank two \\ \hline
		\end{tabular}}
	\end{table}
	
	\begin{table}
		\centering
		\caption{Relationships between tensors}\label{tab16}
		\fontsize{10}{14}\selectfont{
			\begin{tabular}{c||c|c|c|c|c|c} \hline 
				& $A=$ & $A'=$ & $P=$ & $P'=$ & $C=$ & $C'=$\\ \hline \hline
				$A\rightarrow $ & $\times $ & $R A$ & $A R^T $ & $R A R^T $ & $A^T A$ & $R A^T A R^T$\\ \hline
				$A'\rightarrow $ & $R^T A' $ & $\times $ & $(A' )^T $ & $R (A' )^T $ & $(A' )^T A' $ & $R (A' )^T A' R^T$\\ \hline
				$P\rightarrow $ & $P R$ & $P^T$ & $\times $ & $RP$ & $R^T P^T P R$ & $P^T P$\\ \hline
				$P'\rightarrow $ & $R^T P' R$ & $P' R$ & $R^T P' $ & $\times $ & $R^T (P' )^T P' R$ & $(P' )^T P'$\\ \hline
				$C\rightarrow $ & $$ & $$ & $$ & $$ & $\times $ & $R C R^T$\\ \hline
				$C'\rightarrow $ & $$ & $$ & $$ & $$ & $R^T C' R$ & $\times $\\ \hline
				
		\end{tabular}}
	\end{table}

	The coefficient of determination can be considered as a measure of similarity between two random variables. The greater the common variances, measured by the coefficient of determination, the variables are more similar. 
	Table \ref{tab12} contains the coefficients of determination between the primary variables and the principal components. Table \ref{tab14} contains Table \ref{tab12} reduced to the first two rows. By analyzing Table \ref{tab14}, it can be seen that the three primary variables have considerable common variances with the first principal component $p_{c1}$ and one primary variable has a considerable common variance with the second principal component $p_{c2}$. The first, third and fourth primary variables will be in a set of variables similar to the first principal component $p_{c1}$. The second primary variable will remain in a one-element set of primary variables similar to the second principal component $p_{c2}$. This can be seen in Figure \ref{fig5}. Based on this, primary variables can be clustered according to the strength of similarity to the principal components.
	A naive method can be used for clustering.
	If the similarity between a given primary variable and $i-$th principal component is not less than $50\%$, then a given variable belongs to the $i-$th cluster.
	For variables that are not qualified for any cluster, you must create an additional cluster. 
	A more complex method, such as the $k-$means method may also be used.
	For this purpose it can be considered the metric $l_p$ (Euclidean, Manhattan, Chebyshev for $p = 2$, $1$ and $\infty$, respectively) or the cosine similarity \cite{deza2009encyclopedia}. 
	It should be noted that in the case of a greater number of selected principal components, the clusters obtained by different methods may differ.
	
	\section{Discussion}
	Several new results have appeared in this article. The above propositions suggest some unresolved problems that need to be clarified or at least commented on. If this is not possible then problems should be formulated for a later solution.
	\subsection{Virtual representation of primary variables and principal components}
	Real variables Sepal Length, Sepal Width, Petal Length as well as Petal Width are the primary variables. 
	Real variables are obtained from the measurement. 
	For the analysis of the principal components, standardized random variables are considered. 
	Principal components are the same standardized primary variables described in the transformed coordinate system. 
	A single point in a space is the same point, regardless of whether it is described as a point in the standard base, or a point in the base of the eigenvectors.
	Both in the case of standardized primary variables and in the case of principal components we have a real data type.
	
	On the other hand, both primary variables and principal components have their vector representation. 
	In this representation only statistical information is contained. It is information about the variance of individual standardized primary variables or principal components, as well as their interdependence. There is no access to single points in measurements space. 
	It can be said that the vector representation is a representation in the virtual vector space. 
	
	There are some new questions to which the answer is not yet known.
	Can the virtual representation obtained in this work be practically used in any way?
	Can it be useful in data analysis?
	\subsection{Symmetry of matrix $A$ representing primary variables in standard base}
	It was shown that the method of principal components can be represented in a virtual vector space. 
	This applies both to the standardized primary variables and to the principal components. 
	As a result of the calculations, a symmetric matrix (\ref{Eq16}) of vectors representing the primary variables in the standard base was obtained. 
	It can be shown that the result is correct. 
	Left-multiplying the equation (\ref{Eq15}) by a matrix $R$, the equation is obtained:
	
	\begin{equation}\label{Eq20}
	A'=RA.
	\end{equation}
	Using equations (\ref{Eq20}), (\ref{Eq14}) and (\ref{Eq17}) the $P'$ in the following form can be shown:
	\begin{equation}\label{Eq21}
	P'=RA^TR^T.
	\end{equation}
	From the formula (\ref{Eq18}), it can be seen that $P'$ in the base of eigenvectors is a diagonal matrix.
	Thus, there is symmetry:
	\begin{equation}\label{Eq22}
	P'=(P')^T.
	\end{equation}
	By comparing the formula (\ref{Eq21}) with its transposition, the identity is obtained:
	\begin{equation}\label{Eq23}
	A=A^T.
	\end{equation}
	This means that the $A$ matrix representing the standardized primary variables in the standard base is symmetric.
	Therefore, the question arises: Can the information contained in this symmetry be useful in data analysis?	
	\subsection{Tensor Data Mining}
	In the presented analysis, tensor operations were identified.
	These operations consist in finding a tensor description in a rotated coordinate system. 
	The first example is the transformation (\ref{Eq11}) executed on a vector (first order tensor). For a given point described in the standard base, this operation finds its coordinates in the base of the eigenvectors. 
	If this transformation is applied to all points in the space of the standardized primary variables, then all points in the principal component space will be obtained.
	In virtual representation, vectors (first rank tensors) represent the primary variables as well as the principal components.
	On the other hand, tensor of the second rank is the matrix of correlation coefficients and its form after diagonalization. 	 

	Operations on tensors are summarized in Table \ref{tab15}.
	Table \ref{tab16} contains possible to obtain relations between tensors. 
	Six expressions of Table \ref{tab15} can be identified in cells in the Table \ref{tab16} located directly above and below the main diagonal.
	\subsection{Anisotropy of data}
	In the colloquial meaning, anisotropy consists in the fact that some observed quantity depend on the coordinate system. 
	The opposite of anisotropy is isotropy, which lies in the fact that the quantity in all directions is the same.
	Tensors are natural carriers of anisotropy. 
	Since tensor operations have been identified, anisotropy has been identified too.	Because the tensors represent random data, so the data anisotropy was also identified.
	
	Anisotropy of data can be seen in the fact that a single point in the space of real data can be described in different ways. 
	Once it is visible, as a single row in Table \ref{tab3} that contains standardized primary variables. 
	After transformation (\ref{Eq11}) becomes a row in Table \ref{tab9} containing points in the principal components space.
	Anisotropy can also be seen in the fact that the vector representation of the standardized primary variables once takes the form of matrix $A$, another time it takes the form of matrix $A'$. 
	The situation is similar in the case of the representation of principal components, once in the form of the matrix $P$, and again in the form of the matrix $P'$. 	
	The above examples refer to vectors (first rank tensors). 
	
	In turn, symmetric matrix of correlation coefficients (second rank tensor), depending on the coordinate system, is $C$ matrix or $C'$ matrix. 
	In all these cases, the same objects (vectors, second rank tensors) observed in different coordinate systems (different bases) are described by different components.
	They are all anisotropic objects.
	
	\subsection{Clustering of random variables}
	The paper proposes the possibility of clustering correlated random variables because of their similarity to the principal components. It should be noted that clustering of points is commonly used in data analysis. For this purpose many different methods are used, from the classical $k-$means \cite{hand2001} \cite{larose2014discovering}, to the spectral methods \cite{Luxburg2007}. On the other hand, the data table consists of columns and rows. The columns contain random variables and the rows contain points in the space of these variables. It can be said that in the data analysis, the rows in the data table are clustered.
	
	So far, the author has not encountered clustering of random variables (columns in the data table). Perhaps to distinguish between these two types of clustering, it is sensible to use different naming conventions. We suggest that clustering of points in the space of random variables was called horizontal clustering, and clustering of random variables (columns) was called vertical clustering.
		
	\section{Conclusions}
	In this paper the method of principal components was analyzed.
	As a result of this analysis some interesting results have been achieved. 
	These results will be presented synthetically below:	\begin{enumerate}
		\item A geometric analysis of the determination coefficient was performed.
			This analysis led to the conclusion that the coefficient of determination describes the level of common variance of two correlated random variables.
			It was also found that this coefficient is a good measure describing the similarity between the correlated random variables.
		\item Geometric interpretation of the principal component method in a vector space was proposed. 
			This interpretation is based on the generalization of the Pythagorean theorem. 
			For this purpose, correlation coefficients and coefficients of determination between the primary variables and the principal components obtained were analyzed:
			\begin{enumerate}
				\item 	It has been found that standardized primary variables can be decomposed into the sum of independent (orthogonal) random variables. In this way, the variance of each standardized primary variable consists of the sum of the variances of these mutually independent (orthogonal) components. On the one hand it has been noted that the correlation coefficients between the primary variable and the principal components can be interpreted as vector components representing the standardized primary variable in the virtual vector space.On the other hand, the modules of these correlation coefficients can be interpreted as the standard deviation of the above-mentioned independent random variables.
				\item A similar interpretation can be applied to the principal components. Each principal component can be decomposed into the sum of mutually independent (orthogonal) random variables. In the same way, the variance of each principal component consists of the variances of these mutually independent (orthogonal) components. Similarly, as for standardized primary variables, there is also a vector representation for the principal components.
			\end{enumerate}
		\item As a result of the analysis, the principal component method has been enriched with two additional elements:
		\begin{enumerate}
			\item The first element is the proposed algorithm for clustering primary variables by their level of the similarity to the principal components. For the distinction between clustering of points (clustering of rows in a data table) and clustering of variables (clustering columns from a data table), the terms "horizontal clustering" and "vertical clustering" have been proposed. The clustering of the variables proposed in this article is vertical clustering.
			\item The second element is an additional criterion for selection of the correct number of principal components because of the level of reconstruction of the primary variables. Using this criterion will select the appropriate number of principal components in such a way that the level of representation of each primary variable was at least satisfactory.
		\end{enumerate}
		\item A distinction has been made between real data and virtual data. 
			Real data are the primary variables obtained from the measurements as well as the principal components. 
			A vector representation of standardized primary variables and principal components can be called a virtual representation of data or simply virtual data.
		\item Operations on tensors have been identified. In this way it is stated that the principal component analysis is a part 	of the broader concept called Tensor Data Mining.
			Since tensor operations have been identified and tensors are the natural carrier of anisotropy, the concept of Data Anisotropy has been introduced.	
	\end{enumerate}
	\section*{Acknowledgments}
	The term "Anisotropy of Data" was proposed by Professor Micha{\l} Grabowski, when the author of this article stated that in data analysis can be seen tensors as in the case of anisotropic physical phenomena.

	\bibliography{PCA}\label{bibliography}
	\bibliographystyle{unsrt}
\end{document}